\documentclass[12pt,english]{article}
\usepackage{natbib, graphicx, float, subfig, amsmath, bm}

\usepackage{geometry, color, soul}
\usepackage{array}
\usepackage{verbatim}
\usepackage{rotating}
\usepackage{graphicx}

\newcommand\T{\rule{0pt}{3.5ex}}

\begin{document}

\title{Semiparametric Bayesian Density Estimation with Disparate Data Sources: A Meta-Analysis of Global Childhood Undernutrition}
\author{Mariel M. Finucane, Christopher J. Paciorek, \\ Gretchen A. Stevens, Majid Ezzati}
\maketitle

\begin{abstract}
Undernutrition, resulting in restricted growth, and quantified here using height-for-age z-scores,  is an important contributor to childhood morbidity and mortality. Since all levels of mild, moderate and severe undernutrition are of clinical and public health importance, it is of interest to estimate the shape of the z-scores' distributions.

We present a finite normal mixture model that uses data on 4.3 million children to make annual country-specific estimates of these distributions for under-5-year-old children in the world's 141 low- and middle-income countries between 1985 and 2011.   We incorporate both individual-level data when available, as well as aggregated summary statistics from studies whose individual-level data could not be obtained. We place a hierarchical Bayesian probit stick-breaking model on the mixture weights. The model allows for nonlinear changes in time, and it borrows strength in time, in covariates, and within and across regional country clusters to make estimates where data are uncertain, sparse, or missing.

This work addresses three important problems that often arise in the fields of public health surveillance and global health monitoring.  First, data are always incomplete.  Second, different data sources commonly use different reporting metrics.  Last, distributions, and especially their tails, are often of substantive interest. 

\end{abstract}
\thispagestyle{empty}

\newpage

\section{Introduction}
\label{sec:intro3}

Childhood undernutrition contributes substantially to the burden of disease in the low- and middle-income world \citep{black:2008}, and nutritional interventions have the potential to curb child mortality substantially \citep{gakidou:2007}.  Nonetheless, little has been known about the population-level distribution of anthropometric indicators in most countries and regions. 

An important indicator of nutritional status is a child's height-for-age $z$-score ({\sc haz}), which is measured in standard deviations of the height-for-age distribution of the World Health Organization reference population \citep{who:2006}.   Stunting, defined as $\mbox{\sc haz} \leq -2$, and especially severe stunting, defined as $\mbox{\sc haz} \leq -3$, indicate chronic restriction of a child's growth and are associated with adverse physical and intellectual outcomes.  In this paper, we develop a flexible model for the full distribution of {\sc haz} scores that addresses several important challenges.

A barrier to the rigorous population-level analysis of {\sc haz} has been that some surveys report individual-level microdata, whereas others report one or more summary statistics describing their samples, such as average {\sc haz} and/or the sample prevalence of stunting or of severe stunting.  Here we present a novel Bayesian model that combines these disparate data sources. 
 The model incorporates both individual-level data when available, as well as summary statistics from studies whose individual-level data could not be obtained, accounting for dependence when multiple summaries are reported for a single study sample.     

In addition to being able to use data at varying levels of aggregation, we also wish to make coherent inference on both mean {\sc haz} and on all levels of mild, moderate, and severe stunting, since the hazardous effects of undernutrition occur along a continuum.  Exploratory analyses indicated that, across countries and ages, the observed stunting prevalences tend to be somewhat lower than expected under normality.  We concluded that a normal likelihood would not be universally appropriate for our dataset and, furthermore, that assuming normality could potentially induce a systematic bias in estimates of tail probabilities.  Given that stunting prevalences are of substantive interest, this seems especially undesirable.  We therefore developed a single, unified model that allows estimation of all outcomes of interest without assuming normality.   

There is a rich literature on flexible semi- and non-parametric Bayesian methods.  \citet{Dunson:2010} provides a comprehensive overview, focusing on the popular class of Dirichlet process (DP) models.  Many specifications allow for dependence across related DPs \citep[e.g.,][]{MacEachern:1999, DeIorio:2004, Teh:2006, Rodriguez:2008b}.  Semiparametric finite mixture models provide an accurate approximation to the fully nonparametric class of DP models, with sufficient flexibility to fit any smooth density \citep{richardson:1997, roeder:1997}.  Two advantages of the semiparametric approach for our analysis are (a) computational feasibility for very large datasets and (b) avoidance of the `fixed weights' assumption, which can impede efforts to build parsimonious models \citep{chung:2009, rodriguez:2011} and which may be inappropriate for analyses of unbalanced datasets \citep{Venturini:2010}.  

\citet{rodriguez:2009} built from this literature to specify a probit stick-breaking regression model on the weights of a finite normal mixture, keeping the basis distributions constant across groups.  An important contribution of our paper is the extension of their model to the multi-level context.  Whereas \citet{rodriguez:2009} fix certain variance parameters a priori, we estimate all variances, allowing data-driven shrinkage and -- we hypothesize -- inducing an automatic preference for parsimony, discussed in more detail in Section 8.  In addition, we specify a more flexible mean model, allowing for smooth non-linear change over time.  We also extend the  \citet{rodriguez:2009} approach for analyzing microdata to include both microdata and summary statistics in a single model. Although a variety of Bayesian methods for meta-analysis \citep[e.g.,][]{dominici:1999, stangl:2000, higgins:2009} are available, to our knowledge this straightforward approach for combining disparate data sources has not been used previously.  

In this paper, we focus on height-for-age as an example, producing annual estimates of {\sc haz} distributions for under-5-year-old children in 141 low- and middle-income countries from 1985 to 2011.  More generally, variations and trends in health outcomes and risks across the globe have received increased attention in recent years, in part driven by the UN's Millennium Development Goals, the increase in international funding for global health, and the demand for objective evidence about the comparative efficacy of interventions. Our methods are broadly applicable in this arena. For example, this model has been used to analyze weight-for-age $z$-scores and underweight (defined as weight-for-age less than a cutoff) \citep[][]{NIMS:2012} as well as hemoglobin and anemia (defined as hemoglobin less than a cutoff) \citep{stevens2013global}.
 
In the next Section, we describe the {\sc haz} dataset.  Then, in Sec.~\ref{sec:model}, we detail the model's semiparametric likelihood and its hierarchical priors.  Sec.~\ref{sec:computation} describes the  Markov chain Monte Carlo ({\sc mcmc}) algorithm, and Sec.~\ref{sec:results} gives the results of our analysis of undernutrition. We cross validate the model and perform a sensitivity analysis in Sec.~\ref{sec:cv}.  In Sec.~\ref{sec:extensions}, we present two model extensions, and we conclude with a discussion in Sec.~\ref{sec:discussion}.      

\section{The Data}
\label{sec:data}

Our analysis of childhood height-for-age in low- and middle-income countries is based on a vast dataset described in \citet{NIMS:2012}.  The current manuscript presents the statistical methods used in that paper, which was written for a medical audience. The dataset includes 0.7 million individual-level records from health examination, nutrition, and household surveys, as well as summary statistics providing information about an additional 3.6 million children, extracted from the WHO Global Database on Child Growth and Malnutrition and from reports of other national and international agencies. Despite comprehensive access, there are gaps in data availability, with only 126 of the world's 141 low- and middle-income countries contributing data and many countries having data for a limited number of years in the 1985-2011 period.  To facilitate estimation that fills in these gaps, we borrow strength geographically by grouping countries into regions, using the classification described in \citet{NIMS:2012}, and we borrow strength temporally using autoregressive models.  This geographical sparsity is compounded by the fact that less than 85\% of observations are from nationally-representative studies.  Studies that are not nationally representative cannot be relied upon as unbiased estimates of country-level {\sc haz} distributions, because researchers may have tended to select places with systematically more or less severe undernutrition. Furthermore, differences in {\sc haz} distributions across communities in the same country contribute an additional component of variability.  Even a nationally-representative study may not reflect its country's \textsc{haz} distribution with complete accuracy, even apart from sampling uncertainty, due to unobserved study design and measurement issues. On average, the median difference between two national surveys in the same country-year is 0.14 for mean($\mbox{\sc haz}$), 2.47 percentage points for P$(\mbox{\sc haz} \leq -2)$, and 1.82 percentage points for P$(\mbox{\sc haz} \leq -3)$. When all national and subnational surveys are considered, the within-country-year median differences are 0.26, 7.43 percentage points, and 5.96 percentage points, respectively. Yet another source of variability is due to the fact that only 80\% of studies cover the full under-5-year-old age range.  

An important complication is that the data do not come from simple random samples but rather from stratified cluster-based surveys developed for design-based analysis. Weights are assigned to each observation to account for the differences in the probability of being sampled, and these weights can be used to calculate unbiased design-based estimators. In Section 3.5, we describe how we make use of the survey weights and of an estimate of the effective sample size (ESS) of each survey (detailed below) to adjust the likelihood terms in the model (see \citet{chen:2011} for a similar approach). If we did not account for the weights, we could incur serious bias, while if we did not adjust for the ESS, we would mischaracterize uncertainty because of the difference in the amount of information provided by a complex survey compared to a simple random sample. As noted in \citet{gelman:2007statsci}, the appropriate use of data from complex surveys in a modeling context is an area deserving of further research.
 
We estimated the ESS for each survey as the sample size divided by the design effect \citep{Kish:1995}, where the latter is an estimate of the ratio of the variance under the actual design to the variance under a simple random sample \citep[calculated using the \emph{estat effects} command in the Stata 10.1 svy suite,][]{stata:2007}. The design effect can differ for different summary statistics (e.g. mean vs. prevalence below a cutoff), but we need a single design effect for each study since we have a single likelihood contribution for each child's observation, so we calculated an overall design effect as the median of the design effects for the proportion of children whose HAZ score was below ${-4, -3.5, \ldots, 0, \ldots, 3.5, 4}$. The aggregated data sources provide estimates of mean and prevalence that are also based on complex surveys, with the estimates accounting for the survey weights. However the sample sizes reported are the nominal sample sizes and not the effective sample sizes. The design effect can only be estimated from individual-level data though, so we adopted the somewhat ad hoc approach of imputing the ESS for these aggregated data sources by using the median design effect from surveys reporting individual-level data. We used medians rather than means to ensure that an outlying design effect would not overly affect the overall estimate of the design effect. Since the aggregated data sources tend to be large, design effects are imputed for studies amounting to 84\% of the total number of children (3.6 million / 4.3 million), and we acknowledge this as a limitation of our analysis. We note, however, that 59\% of these (2.1 million / 3.6 million) are from a series of four very large surveillance studies in Chile.  In addition, sampling uncertainty represents only a modest proportion of the overall measurement error variability for most data points because of the studies' relatively large sample sizes and the importance of the study-specific random effects (Section \ref{subsec:error}), so misestimation of sampling variance is not likely to have major impact on our inference.

\section{The Model}
\label{sec:model}

In this Section, we present a model that combines information from disparate and sparse data sources to estimate {\sc haz} distributions for each country-year.  After describing the model for the population-level {\sc haz} distribution in each country-year and the hierarchical structure that borrows strength across countries and years, we describe how both individual-level data and aggregated summary statistics, such as sample means and prevalences, are included in the likelihood, in both cases accounting for complex survey design.

\subsection{Semiparametric density estimation for dependent random distributions}
\label{subsec:npBayes}

Let $f_i(z)$ denote the {\sc haz} distribution in study $i$ ($i=1,\cdots,n$) at time $t[i]$ in country $j[i]$ ($j=1,\cdots,J$), with square brackets used to denote group membership.  Building on \citet{rodriguez:2009}, we set:
\begin{eqnarray}
f_{i}(z) &=& \sum_{m=1}^{M+1} w_{mi}\ \mathcal N\left(z\, |\, \theta_m, \sigma^2_m\right) \label{eq:mix}\\ 
\mbox{} \nonumber \\
w_{mi} &=& \left\{ \begin{array}{ll} \Phi(\alpha_{mi}) \prod_{u=1}^{m-1}(1-\Phi(\alpha_{ui}))  & \mbox{if } m\leq M \\ 
\prod_{u=1}^{M} (1-\Phi(\alpha_{ui}))& \mbox{if } m=M+1
\end{array} \right. \label{eq:probit}\\ 
\mbox{} \nonumber \\
 \alpha_{mi} &=& \delta^c_{mj[i]} + \varphi^c_{mj[i]} t[i] +
  u_{mj[i]t[i]} +  \bm{\beta}^{\prime}_m \bm{x}_{j[i]t[i]} + a_{mi} + b_{mi}.  \label{eq:anova}
\end{eqnarray}
Eq.~(\ref{eq:mix}) describes a finite mixture of $M+1$ normal distributions, where the weights ($w$) on the constituent normal distributions vary across studies.  We assign uniform priors to the $\theta_m$'s and $\sigma_m$'s \citep{Gelman:2006}, and we constrain the $\sigma_m$'s to be less than two times the standard deviation across all observed {\sc haz} values.  This constraint is placed to ensure propriety and is chosen to be noninformative. We constrain $\theta_1 < \theta_2 < \cdots < \theta_{M+1}$  to ensure interpretability of the main effects and so that the interpretation of the $\alpha$'s doesn't change across iterations of the {\sc mcmc} chains.

We specify a probit stick-breaking model for the $w$'s in (\ref{eq:probit}).  This transformation uses the standard normal cumulative distribution function ($\Phi$) to transform $\alpha$'s that range between -$\infty$ and $\infty$ to $w$'s that range between 0 and 1.  Specifically, the $\alpha$'s determine the relative weights assigned to each cluster in the following manner: starting with a `stick' of length one, $\Phi(\alpha_{1i})$ is the proportion of the stick that we break off and assign to $w_{1i}$; $\Phi(\alpha_{2i})$ is the proportion of the remaining stick of length ($1-w_{1i}$) that we break off and allocate to $w_{2i}$; and so on.  Larger values of $\alpha_{mi}$ thus correspond to higher weights on the $m^{\mbox{\footnotesize th}}$ mixture component in study $i$.  The probit stick-breaking transformation allows us to place a flexible model on the $\alpha$'s, while ensuring that the $w$'s still add to one.  As an aside, note that although stick-breaking is widely used, it does have an interpretational drawback in analyses such as this one. The stick-breaking prior tends to put the largest weight on the first mixture component, the second-largest weight on the second mixture component, and so forth.  In our model, since we constrain $\theta_1 < \theta_2 < \cdots < \theta_{M+1}$, this tendency would result in a right-skewed distribution, which is not something we'd expect in modeling densities of \textsc{haz}.  The structure of our model allows for non-skewed fits, but an alternative approach would be the additive log-ratio transformation \citep[][p. 113]{Aitchison:1986}.

In  (\ref{eq:anova}), we define $\alpha_{mi}$ to be comprised of six effects that allow us to leverage all available information in making estimates for each country-year pair. $\delta_{mj}^c$ is a component-specific country-level intercept, determining the baseline weight placed on each of the $M+1$ normal distributions in country $j$. $\varphi^c_{mj}$ is a country- and component-specific linear time effect, determining the linear parts of country $j$'s time trend.  The hierarchical priors for the $\delta^c$ and $\varphi^c$ terms are described in Sec.~\ref{subsec:hierarchical}.  Letting $T=27$ be the total number of time-points of interest ($1985, 1986, \cdots, 2011$), the $T$-vector $\bm{u}_{mj}$ captures smooth nonlinear change over time in country $j$ and mixture component $m$, with details provided in Sec.~\ref{subsec:gaussian}.  $\bm{\beta}_m$ contains the effects for mixture component $m$ of time-varying country-level covariates \citep[maternal education, national income, urbanization, and an aggregate metric of access to basic healthcare, as described in][]{NIMS:2012}. The $a$'s are study-specific random effects, and the $b$'s capture the extra variance of studies that do not fully cover the under-5-year-old age range.  These error terms and their variances are described in Sec.~\ref{subsec:error}. 

For purposes of identifiability, it is important to note that each set of random effects captures a distinct dimension of variability in the data.  To be specific, consider Equation \ref{eq:anova}.
The $\delta$'s are intercepts; the $\varphi$'s are linear time slopes (with a zero overall mean); the $u$'s capture nonlinear variation in time (and are constrained to have a zero overall mean and a linear time slope of zero); and the $\beta$'s are covariate effects.  The $a$'s and $b$'s are just auxiliary variables that allow for heteroskedasticity and correlation -- if desired, we could have integrated them out to explicitly make them part of the variance.

\subsection{Hierarchical intercepts and linear time slopes}
\label{subsec:hierarchical}

The model has a hierarchical structure: studies are nested in countries, which are nested in regions (indexed by $k$), which are, of course, all nested in the globe.  For each mixture component, we specify hierarchical priors for the country-specific random intercepts ($\delta^c$) and the country-specific linear time slopes $(\varphi^c)$, with country-specific effects centered around their region-level counterparts ($\delta^r$ and $\varphi^r$) and region-level effects centered on the global means ($\delta^g$ and $\varphi^g$).  For both the intercepts and the time slopes, we use a fat-tailed $t_4$ prior \citep[][p. 451]{Gelman:2004} at the country level to allow for the fact that some countries may deviate substantially from their expected values given covariates and geographic hierarchy:
\begin{center} \begin{tabular}{ll}
$\delta^c_{mj} \sim t_4 \left(\delta^r_{mk[j]}, \tau^{2\,\delta^c}_m \right)$, & $\delta^r_{mk} \sim \mathcal N \left(\delta^g_{m}, \tau^{2\,\delta^r}_m \right)$,\\
$\varphi^c_{mj} \sim t_4 \left(\varphi^r_{mk[j]}, \tau^{2\,\varphi^c}_m \right)$, & $\varphi^r_{mk} \sim \mathcal N \left(\varphi^g_{m}, \tau^{2\,\varphi^r}_m \right).$
\end{tabular} \end{center}
The $\tau$'s determine the degree of intercept ($\tau^\delta$) and slope ($\tau^\varphi$) shrinkage performed at the country- ($\tau^{\delta^c}$ and $\tau^{\varphi^c}$), and region-levels ($\tau^{\delta^r}$ and $\tau^{\varphi^r}$). Whereas \citet{rodriguez:2009} set each $\tau$ parameter equal to 1, we assign a uniform hyperprior to each of the $\tau$'s \citep{Gelman:2006}, truncating such that $\tau>.00001$ (as smaller values are not identifiable from one another).  An advantage of our approach is that the data are able to inform estimation of the $\tau$'s.  This allows for different degrees of shrinkage depending on the signal-to-noise ratio in the data.  We borrow strength across units, compromising between the overly-noisy unit-specific estimate and the overly-simplified all-unit estimate \citep{Gelman:2007}. 

Importantly, we hypothesize that our specification favors a simplified model.  In those cases when the data do not contain evidence to the contrary, we believe that the natural Bayesian penalty on model complexity \citep[which has been shown explicitly in much simpler models,][]{jefferys:1992} causes extraneous effects to be shrunk toward zero, allowing the model to collapse down toward a simplified specification.  This is an especially desirable characteristic given that our model contains such a large number of random effects.  We present evidence to this effect in the Discussion.

\subsection{Nonlinear change in time}
\label{subsec:gaussian}
\textsc{haz} distributions are expected to change smoothly over time.  In country $j$ and mixture component $m$, we capture smooth nonlinear change in time using the $T$-vector $\bm{u}_{mj}$.  We model $\bm{u}_{mj}$ hierarchically by defining it as the sum of country, region, and global components:  $\bm{u}_{mj} = \bm{u}^c_{mj} + \bm{u}^r_{mk[j]} + \bm{u}^g_m.$ In order to allow the model to differentiate among the degrees of nonlinearity that exist at the country, region, and global levels, the $T$-vectors $\bm{u}^c_{mj}$, $\bm{u}^r_{mk}$, and $\bm{u}^g_m$ are each given a second-order Gaussian autoregressive prior \citep{ Breslow:1993, Rue:2005} with mean zero and precision matrices  $\lambda^c_m\bm{P}$, $\lambda^r_m\bm{P}$, and $\lambda^g_m\bm{P}$ respectively. The scaled precision matrix, $\bm{P}$, penalizes second differences.  

For each of the precision parameters, we use a truncated flat prior on the standard deviation scale ($1/\sqrt \lambda$) as recommended by \citet{Gelman:2006}.  We truncate these priors such that $\mbox{log} \lambda \leq 12$ for each $\lambda$.  This upper bound is enforced as a computational convenience: models with $\mbox{log} \lambda > 12$ are considered to be equivalent to a model with $\mbox{log} \lambda = 12$ as they have essentially no extra-linear variability (a.k.a. `wiggliness') in time.  Furthermore, we order the $\lambda$'s a priori: $\lambda_m^c < \lambda_m^r < \lambda_m^g$ for $m=1,\cdots, M$.  This prior constraint conveys the natural expectation that, for example, the global {\sc haz} trend has less extra-linear variability than the trend of any given region.  

The matrix $\bm{P}$ has rank $T-2$, corresponding to a flat, improper prior on the mean and time slope of the $u$'s 
\citep[p.~191]{Wood:2006}.  Thus we have a proper prior in a reduced-dimensional space, $P(\bm{u}|\lambda) \propto \lambda^{\frac{T-2}{2}} \exp\left(-\frac{\lambda}{2} \bm{u}^{\prime} \bm{P} \bm{u} \right)$ \citep{Rue:2005}. 
We constrain the mean and slope of each $\bm{u}$-vector to be zero to avoid non-identifiability with the $\delta$ and $\varphi$ terms. 

 
\subsection{Error terms}
\label{subsec:error}
Nationally representative sources have errors larger than their sampling error, due to unobserved study design and measurement issues (see, for example, the 2002 study from China in Fig.~\ref{fig:countryResults}). The study-specific random effect, $a_{mi}$, allows mixture component $m$ in study $i$ to have an unusually high or an unusually low mixture weight after accounting for the other terms in the model.  We assign independent  $t$ priors to the $a$'s to account for the fact that some unusual studies may deviate substantially from their expected country-year mean: $a_{mi} \sim t_4\left(0, v_{mi}\right)$. The scale parameter, $v_{mi}$, depends on whether the study is national or subnational:
\begin{eqnarray*}
v_{mi}= \left\{ \begin{array}{ll}
                    v_m^n & \mbox{if study $i$ is nationally representative} \\
                    v_m^s& \mbox{otherwise.} \end{array} \right.
\end{eqnarray*}
Random effects from nationally-representative studies are constrained to have less variability than random effects from other studies ($v_m^n < v_m^s$ for $m = 1, \cdots, M$). Thus the random effects for non-nationally-representative studies account for both the non-sampling errors mentioned above and for within-country heterogeneity.

As an aside, we note that a limitation of our model is that we may obtain biased estimates in a country with many non-representative studies that share a common, systematic bias. One possible path for future research to address this issue would be an overall bias term, estimated from non-representative data across multiple countries, or bias terms for each country, with an overall shrinkage prior.  

For studies that do not fully cover the under-5-year-old age range, we include an additional error term for each mixture component to capture variability of {\sc haz} distributions across ages: $b_{mi} \sim t_4(0, v^b_m)$.  For studies that do cover the full age range, we set $b_{mi} = 0$ for $m=1,\cdots, M$.

Note that to make country-level predictions, we set $a = b = 0$, thus removing the random effects due to imperfections in study design and to within-country and across-age variability of {\sc haz} distributions.

\subsection{The likelihood, accounting for complex survey design}

In the likelihood terms that relate the data to the population-level distribution in a given country-year (Eq. \ref{eq:mix}), we account for both the survey weights and the survey effective sample size (ESS). 

\subsubsection{Likelihood for individual-level data}

Our initial strategy was to take a weighted (using the survey weights) resample from the observations in a given survey, with the number of samples equal to the ESS for the survey; this strategy was used for our primary analysis \citep[Section~\ref{sec:model} and][]{NIMS:2012}, with the likelihood for a given survey being the product across the resampled observations. 

However, resampling introduces an additional component of variability, the randomness in the realized resample. For the analysis described in Section~\ref{subsec:mom} and in \citep{PaciorekLGH}, we developed a new approach of using a weighted likelihood, normalizing the survey weights to sum to the survey ESS. We used the survey weight for each observation as the weight in the likelihood to account for the sampling design, and we adjusted the weights in a survey such that taken together the weights sum to ESS and not the nominal sample size. Thus, the contribution to the log-likelihood for an individual was the normalized weight for that individual multiplied by the log of the relevant country-year density (Eq. \ref{eq:mix}). Note that the analyses that used the weighted likelihood were also analyses that produced separate estimates for urban vs. rural place of residence. Because the surveys generally stratified by urban/rural residence and were designed to provide valid estimates for both strata, we estimated ESS and scaled the weights separately for urban and rural data. 

\subsubsection{Likelihood for aggregated data}
\label{subsec:grouped}

Recall that a motivating goal of this analysis is to specify a model that incorporates both individual-level and aggregated data.  Here we describe how we included summary statistics, accounting for dependence when multiple summaries are reported for a single sample, as well as accounting for complex survey design. In the case of our example analysis, including summary statistics enables us to include data from 349 studies for which we were unable to obtain microdata.  The sample sizes are large, ranging from 200 to 643,200 children, with a median sample size of 2,000.  

For each study, we have one or more of the sample mean $\mbox{\sc haz}$, sample prevalence of stunting ($\mbox{\sc haz} \leq -2$), and sample prevalence of severe stunting ($\mbox{\sc haz} \leq -3$), where the summary statistic is a design-based estimator that accounts for the survey weights. We accounted for dependence across multiple summaries reported for a single study sample, e.g., a sample's mean and one or more of its sample prevalences as follows. If the data were a simple random sample, then each of the summary statistics could be written as the sum of a large number of independent random variables. Based on the multivariate central limit theorem, we derived the resulting multivariate joint normal likelihood for the summary statistics of each sample based on the population distribution (Eq. \ref{eq:mix}, see Appendix for technical details). Of course the summary statistics are also based on complex survey designs, so we used the imputed ESS in place of the actual sample size in the multivariate likelihood for each study, thereby giving us an approximate likelihood that treats the design-based estimates as if they came from a simple random sample with an adjusted sample size to reflect the change in information available from a complex survey. \citet{chen:2011} developed a similar approach independently.

We conduct a simulation study to determine whether the aggregated-data sample sizes are large enough for multivariate normality to be a reasonable approximation of the data distribution.  Consider a study that reports microdata.  We simulate the joint sampling distribution of its three summary statistics as follows.  We first obtain a rough estimate of the {\sc haz} density for that study using a kernel density estimator.
We then take 1000 draws of datasets of size $n$ from the estimated density to simulate the unobserved individual-level {\sc haz} values underlying a given set of summary statistics.  We take $n=199$, the smallest observed aggregated-data sample size, since our aim is to ensure that joint normality holds for all studies.  We summarize each simulated dataset to obtain 1000 draws of mean({\sc haz}), P($\mbox{\sc haz} \leq -2$), and P($\mbox{\sc haz} \leq -3$).  

\begin{figure}[!t]
\centering
\subfloat[\emph{When the prevalence of stunting is large, the normal likelihood provides a good approximation to the true sampling distribution of summary statistics.}]{
\includegraphics[width=2.8in]{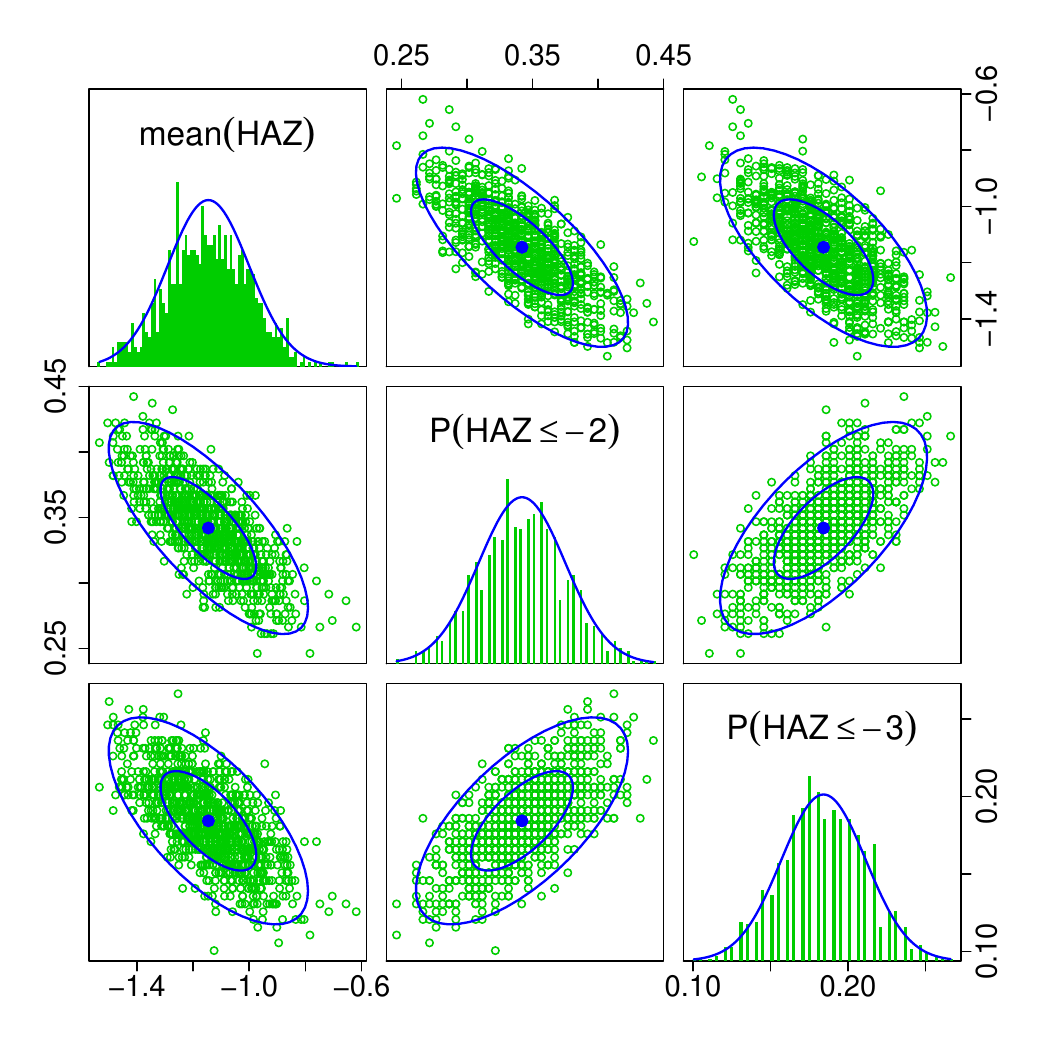}
}\qquad \hspace{-.25in}
\subfloat[\emph{Even when the prevalence of stunting is small, the normal likelihood still provides  a reasonable approximation to the the true sampling distribution.}]{
\includegraphics[width=2.8in]{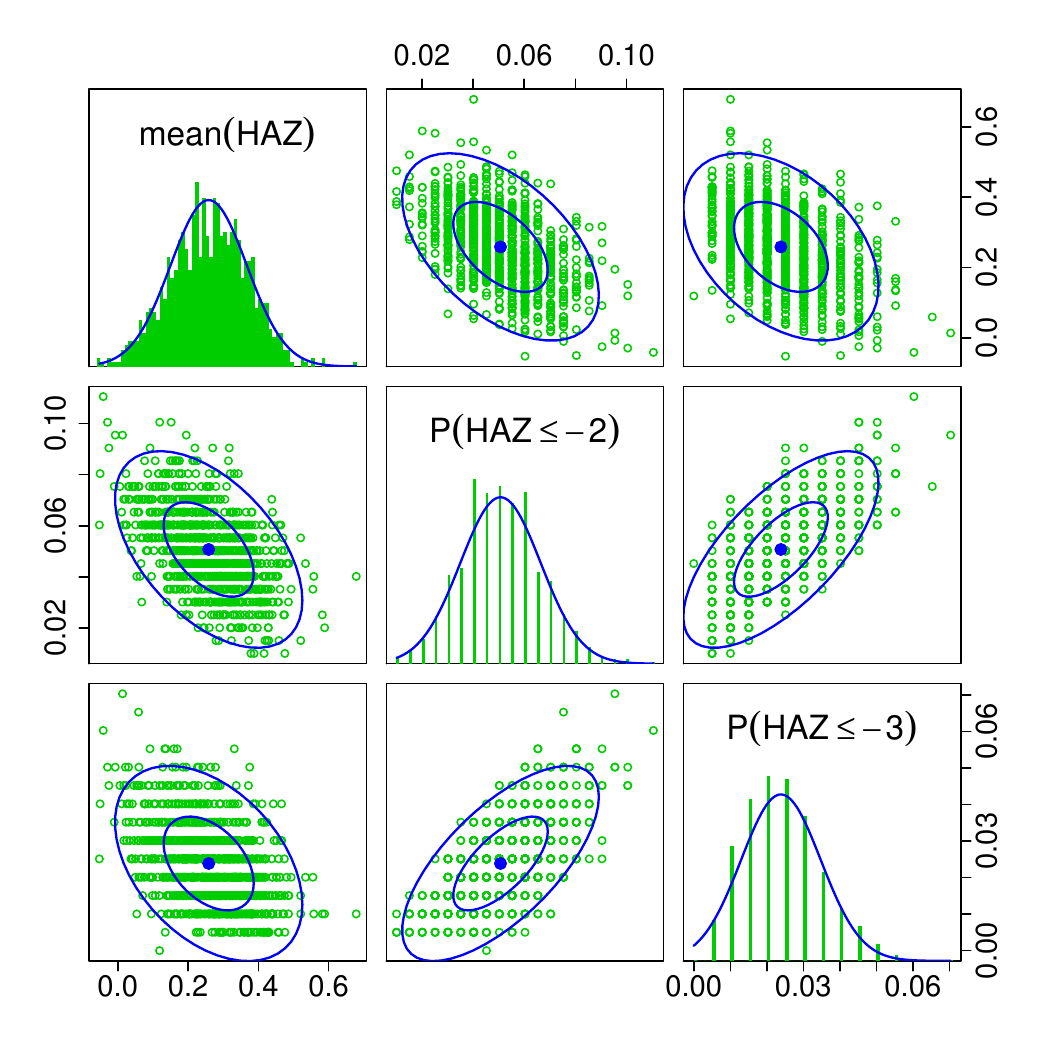}
}
\caption{In green, we show univariate and bivariate projections of the simulated three-dimensional joint sampling distribution of  aggregated-data summary statistics given a sample size of 199.  In blue, we overlay the assumed normal distributions.}
\label{fig:CltCheck}
\end{figure}

We repeat this simulation for each study that reports microdata.  In Fig. \ref{fig:CltCheck} we show results for two example studies  -- one with a large stunting prevalence and one with a small stunting prevalence.  We see that even for an $n$ as small as 199, and even when stunting prevalences are low, the true sampling distribution of the summary statistics is still a reasonable approximation to the joint normal likelihood.  For aggregated-data studies with larger samples sizes, the normal assumption provides an even better approximation (results not shown).

To obtain the full likelihood, allowing coherent posterior inference that incorporates data reported at both levels of aggregation, we simply multiply the microdata likelihood with the aggregated-data likelihood.

\section{Computation}
\label{sec:computation}

We fit the model via {\sc mcmc}, programming the sampler using the statistical computing language \verb~R~ \citep{R:2008}.  Algorithms to fit mixture models often introduce latent variables that break the mixture and thus enable Gibbs sampling \citep{rodriguez:2009}.  However, since our dataset includes summary statistics,  we would have had to impute the  individual-level values underlying each summary statistic in order to allow Gibbs sampling.  Given the large sample sizes and the fact that the imputed samples would have had to satisfy data-imposed constraints on their means and tail probabilities, this option was not feasible.  

As an alternative, we considered a Metropolis-Hastings algorithm that breaks the mixture for the microdata and then proposes model parameters from their closed-form full-conditional distributions given the microdata only.  In such a high-dimensional model though, good mixing relies heavily on an algorithm that makes moves with a covariance roughly proportional to the full posterior covariance.  Since these proposal covariances did not take the aggregated data into account, mixing under this algorithm was very poor.

In the end, we therefore opted for a Metropolis-within-Gibbs algorithm.  We used an adaptive Metropolis algorithm due to \citet{shaby:2011} to update the $\alpha$'s, and we used Gibbs sampling for the parameters in the hierarchy.  Note that a key strategy to achieve better mixing was to jointly sample $\bm{\theta}$ and $\bm{\sigma}$, the means and variances of the normal mixture components, because of the strong  dependence across these parameters.  Also, for each $m$, we jointly sampled $\lambda_m^c$ with $\{\bm{u}_{mj}^c\}_{j=1,\ldots,J}$, $\lambda_m^r$ with $\{\bm{u}_{mk}^r\}_{k=1,\ldots,K}$, and $\lambda_m^g$ with $\bm{u}_{m}^g$, because of the strong dependence between the $\lambda$'s and $\bm u$'s. We did not marginalize over mean parameters in the model, despite having some normal conditional distributions, because this would have introduced off-diagonal structure into the likelihood covariance, requiring manipulation of large covariance matrices to calculate the marginal densities. 

We started five chains in parallel at randomly-selected starting values. We phased  blocks of parameters  into the \textsc{mcmc} sequentially, allowing each block to adjust to the data before constraining it by starting the next block.  In particular, we started the $\alpha$'s first, holding all other parameters constant at their starting values.  This allowed the $\alpha$'s to reach reasonable `pseudo-data' values before we began updating the $\theta$'s, $\sigma$'s, and the country-level effects. 

In order to obtain a sufficient number of effectively independent samples from the posterior, we ran 5 chains for 30,000 iterations after 22,000 iterations of burnin.  We then combined the five chains and thinned by a factor of 30 to obtain chains of length 5,000 with which to summarize the posterior.  We obtained convergence of the chains for mean({\sc haz}) and P($\mbox{\sc haz} \leq -2$), with Gelman-Rubin statistics \citeyearpar{gelman:1992} of $R<1.1$ for all country-years.  For P($\mbox{\sc haz} \leq -3$), there were 9 country-years (5 from Kuwait and 4 from Venezuela) with $1.1 < R < 1.2$.  In addition, for all country-years, all three metrics had posterior standard deviations at least 10 times greater than their Monte Carlo standard error. Due to the lack of identifiability inherent in mixture likelihoods, the $\theta$ and $\sigma$ parameters mixed least well -- we obtained only 23--208 effectively-independent samples of these parameters. The imperfections in mixing of the mixture component parameters did not impact mixing at the level of inference however. The country-year {\sc haz} means, for example, had an average of 3962 effectively-independent samples.

\section{Results}
\label{sec:results}
\begin{figure} 
\centering
\includegraphics[width=5in]{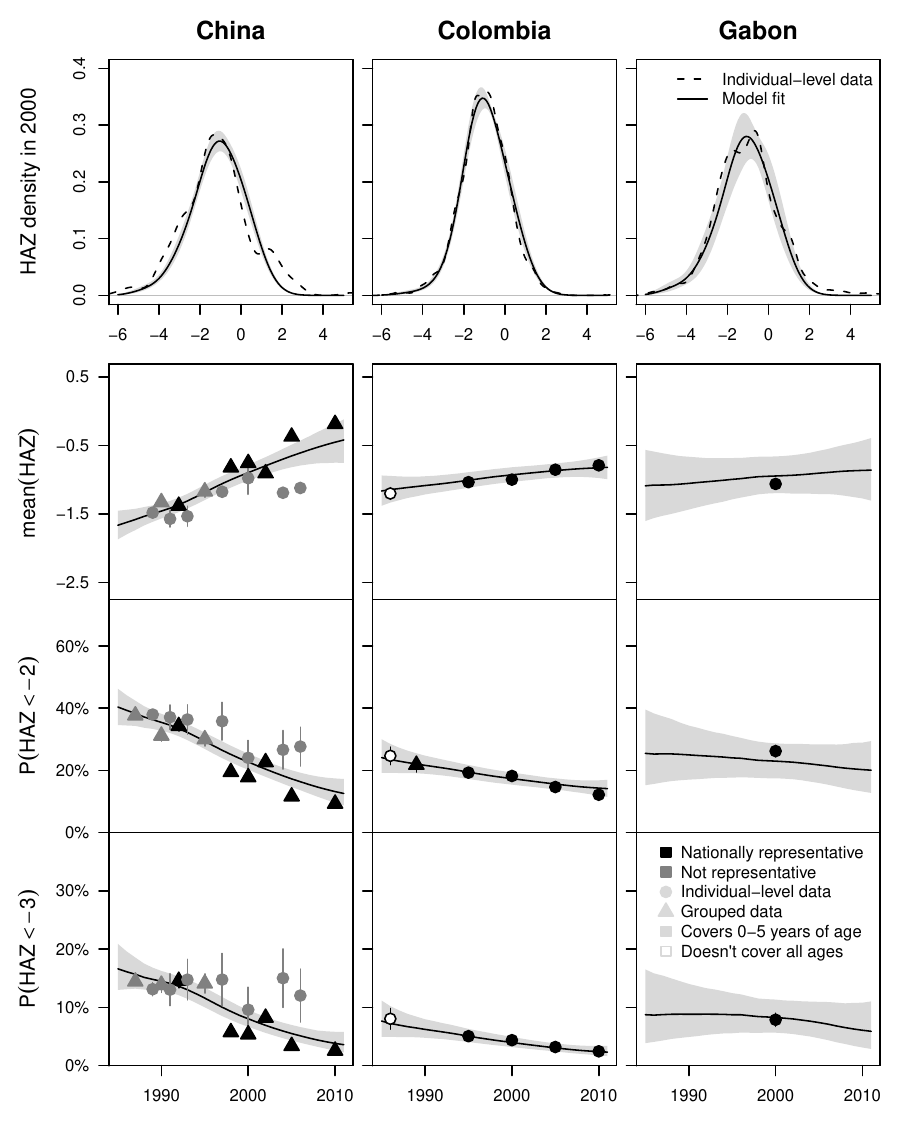}
\caption{Three examples of country-level results.  The estimated {\sc haz} density in the year 2000 is shown in the first row as a solid black line, with pointwise 95\% uncertainty bands in grey.  The raw data is estimated with a kernel density estimator (black dashed line) for comparison. The last three rows show posterior mean trend estimates (black) and their 95\% credible intervals (grey) for three summary statistics of interest. Data points are indicated by symbols, with 95\% uncertainty intervals reflecting sampling variability.}
\label{fig:countryResults}
\end{figure}

We find that across age groups and countries, {\sc haz} distributions are unimodal and roughly symmetric (see Fig.~\ref{fig:countryResults} for a few examples).  As a result, ($M$+1)=5 clusters provide sufficient flexibility to fit these data \citep{richardson:1997}.  We present a sensitivity analysis of our results to the value of $M$ in Sec.~\ref{sec:cv}.  

In the first row of Fig.~\ref{fig:countryResults}, we demonstrate how the semiparametric weighted-mixture model fits three example {\sc haz} densities.  Note that in Colombia and Gabon, where data are richer (with $n$=3,221 and $n$=2,251 individual-level year-2000 observations, respectively), the modeled densities follow the data-based empirical densities more closely than in China, where data are relatively sparse ($n$=227 individual-level year-2000 observations).  In Colombia, where the density is tighter, fewer component normal distributions are given non-negligible weight relative to China and Gabon, where the densities are more diffuse. 

Similar lessons can be drawn by examining the model's fit to the summary statistics shown in the last three rows of Fig.~\ref{fig:countryResults}.  We achieve narrower confidence bands in data-rich China (total $n$=111,471 across all data sources) and Colombia (total $n$=23,680) than in relatively data-poor Gabon (total $n$=2,251).  Despite Gabon's data sparsity, we are nonetheless able to make reasonably certain inference, with estimates informed by the rich data available from other countries in Sub-Saharan Africa and, to a lesser extent, by covariates and data from other regions as well.  The uncertainty of these estimates reflects the degree of country-to-country and region-to-region variability observed in 
other parts of the dataset.

\begin{figure}
\centering
\includegraphics[width=6in]{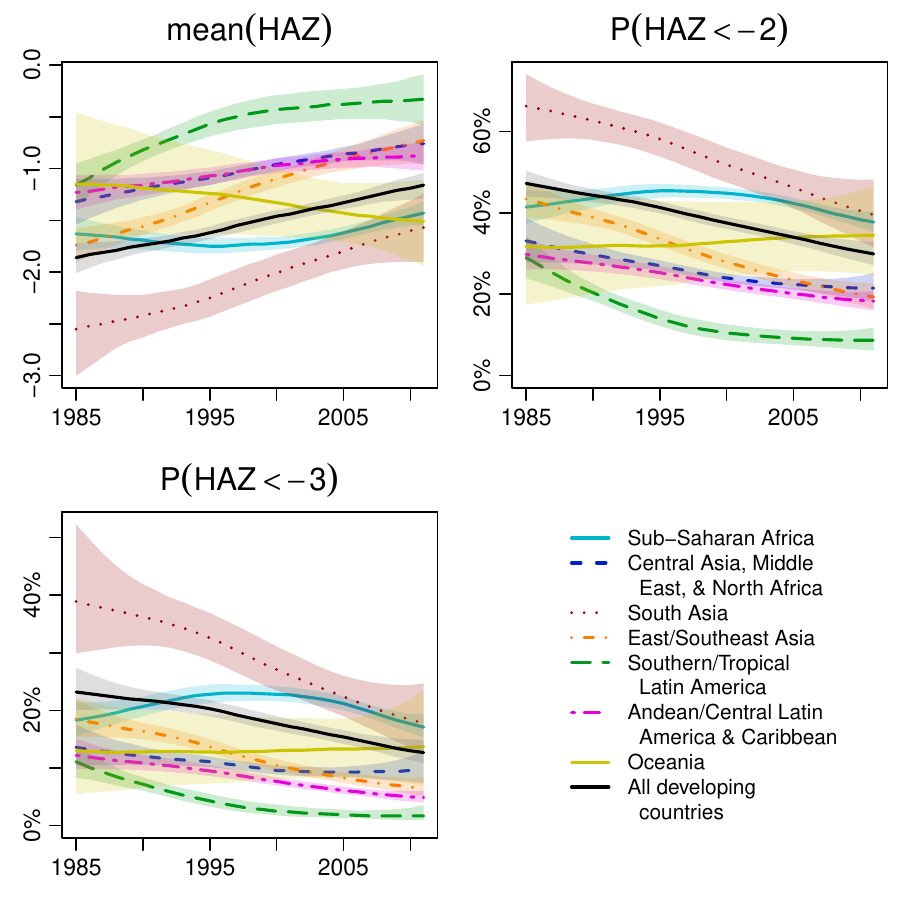}
\caption{Population-weighted region- and global-level trend estimates (solid colors) and their pointwise 95\% uncertainty bands (transparent colors).}
\label{fig:sregionFits}
\end{figure}

A motivating goal of this analysis is to estimate population-level distributions of {\sc haz}, not just at the the country level, but also for regions and for the globe as a whole.  In Fig.~\ref{fig:sregionFits}, 
we show region- and global-level results.  These estimates are calculated using weighted averages of each region's constituent country-level distributions, weighting by the size of the country's population.  All 141 of the world's low- and middle-income countries contribute to these calculations, including the 15 countries that have no data.  This allows us to reflect our uncertainty due to data sparsity, in contrast to models that deterministically set health metric levels in countries without data to some pre-specified value \citep[e.g.,][]{mason:2005}. 

Substantive findings on undernutrition in children are discussed in full in \citet{NIMS:2012}.  In short, we find that although anthropometric status in developing countries has improved on average between 1985 and 2011, worldwide progress has been uneven.  The largest absolute improvements in \textsc{haz} means occurred in Asia and the largest relative reductions in prevalence in Southern and Tropical Latin America. Anthropometric status worsened slightly in sub-Saharan Africa from the mid 1980s to the late 1990s, when it began improving. In 2011, 314 million (95\% uncertainty interval 296-331 million) under-five children were mildly, moderately or severely stunted. Most of these children lived in South Asia and sub-Saharan Africa. 

\section{Model checking}
\label{sec:cv}

Out-of-sample performance is of particular relevance in this analysis because (a) such a complex and highly-structured model is at risk for overfitting, and (b) one of the primary purposes of our modeling efforts is to make predictions and report uncertainties for country-years without data.  For these reasons, we conducted an extensive cross-validation of our model as well as two other candidate models.

Our first alternative candidate model was actually a trio of models that made separate estimates for means and for each of the two prevalences without sharing information across them in the form of a distribution.  Microdata were summarized and combined with grouped data.  Means and prevalences (using a probit regression formulation) were each modeled separately using an identical hierarchical structure to the main model in terms of covariate effects, shrinkage priors, non-linear time trends, etc.  These separate models for each reporting metric had the advantage of mixing quickly and being easy to describe to global health researchers.  Due to concern about the predictive power of our covariates, our second alternative candidate model was the same as the primary model but excluded all covariates.

Our five-fold cross validation consisted of two tests. In Test 1, in each of the five folds, we held out all data from non-overlapping subsets of 10\% of countries with data (i.e., created the appearance of countries with no data where we actually had data), with held-out data from a mix of countries that were data rich ($\geq$ 6 years of data with at least one year of data after 1999), data poor ($\leq$ 3 years of data), and that had average data density (4-5 years or $\geq$ 6 years of data with no data after 1999). We fit the model to the data from the remaining 90\% of countries and made predictions of the held-out observations. We calculated the absolute and relative differences between our predictions and the held-out data. 

The hold-out algorithm for Test 2 was designed to examine how well the model predicted mean and prevalence $<$ -3 when only prevalence $<$ -2 was available, as commonly occurred.  (See, for example, the 1989 study from Colombia, in Fig.~\ref{fig:countryResults}.) For  countries with data on mean as well as prevalences below -2 and -3, we maintained prevalence $<$ -2 and held out one or both of the other two metrics, such that data from a total of 10\% of all studies were excluded. We fit the model to the remaining 90\% of the dataset and made predictions of the held-out observations.

As shown in Table \ref{tab:cv}, our main model (M1) predicted the known-but-masked truths well -- across metrics, the median absolute prediction error (0.21, 6.02, and 3.62 for mean, \%$\leq$-2, and \%$\leq$-3, respectively) is smaller than the median difference between two surveys in the same country-year (0.26, 7.43, and 5.96, respectively). 

\begin{table} 
\begin{center}
\begin{tabular}{lc||c|cc|cc||c|cc|cc}
\hline
&& \multicolumn{5}{c||}{\textbf{Median absolute error}}& \multicolumn{5}{c}{\textbf{Median relative error}}\\
 & $n$ & M1 & M2 & (p) & M3 & (p) & M1 & M2 & (p) & M3 & (p) \\ 
  \hline
  \multicolumn{12}{l}{\textbf{\underline{Test 1: Holding out all studies from 10\% of countries}}} \T \\
mean \textsc{haz} & 181 & 0.21 & 0.20 & 0.16 & 0.28 & 0.00 & 0.15 & 0.14 & 0.56 & 0.19 & 0.00 \\ 
  \%\textsc{haz}$<$-2 & 292 & 6.02 & 5.94 & 0.36 & 8.05 & 0.00 & 0.17 & 0.17 & 0.19 & 0.25 & 0.00 \\ 
  \%\textsc{haz}$<$-3 & 202 & 3.62 & 3.40 & 0.57 & 4.12 & 0.00 & 0.27 & 0.26 & 0.26 & 0.30 & 0.00 \\ 
  \multicolumn{12}{l}{\textbf{\underline{Test 2: Holding out mean and \%$<$-3 when \%$<$-2 is known}}} \T \\
mean \textsc{haz} & 83 & 0.10 & 0.16 & 0.00 & 0.10 & 0.72 & 0.10 & 0.15 & 0.00 & 0.10 & 0.89 \\ 
 \%\textsc{haz}$<$-3 & 111 & 1.76 & 1.86 & 0.00 & 1.76 & 0.19 & 0.16 & 0.19 & 0.01 & 0.16 & 0.30 \\ 
   \hline
\end{tabular}
\caption{Errors in predictions of held-out data for three candidate models. M1 is the main model; M2 is a trio of separate models for mean and for each of the two prevalences; M3 excludes all covariates. $n$ gives the number of held-out observations. Prediction errors were compared across models using the non-parametric Wilcoxon signed-rank test for paired data \citep[as discussed in][]{dietterich:1998}, with p-values in the (p) columns for the null hypotheses that the median differences between the errors of M2 and M3 with M1 are zero. The tests were conducted assuming independence of the held-out values.  The p-values should therefore be interpreted as an approximation because there is some dependence among the held-out observations.
}
\label{tab:cv}
\end{center}
\end{table}
Furthermore, our model made better out-of-sample predictions than the model without covariates (M3) when we held out all studies from a given country (Test 1).  By contrast, our model  did not make better out-of-sample predictions than the trio of separate models (M2) when we held out all studies from a given country (Test 1).  We hypothesize that the predictive accuracy for a new country for a given metric of interest is robust to the differences between M1 and M2 because these predictions rely primarily on
the region-level estimates for the metric, which are likely driven primarily by data on the metric of interest 
rather than by data on other metrics.  We acknowledge that Table 1 makes it appear that M2 in fact has a slight edge on M1, but we emphasize that this edge is not statistically significant -- we believe this is due to the chance in our random selection of data points to hold out in the cross validation.  

Importantly, the main model had lower error than the trio of separate models in predicting held-out mean and prevalence $<$ -3 (the more-commonly missing indicators) when prevalence below -2 was observed (Test 2). This advantage arises because, by estimating the whole distribution, the mixture model leverages the information contained in one metric to make estimates for another metric that was not observed.  Using the year 2000 as an example, we note that there is a high degree of correlation across countries of the posterior means of mean({\sc haz}), P($\mbox{\sc haz} \leq -2$), and P($\mbox{\sc haz} \leq -3$),  likely due at least in part to constraints on the possible values, with all correlations exceeding 0.95 in absolute value. 

In Test 2 the main model had identical errors to the model without covariates, suggesting that conditional on P($\mbox{\sc haz} \leq -2$), the covariates provide no additional information about the unobserved values of mean({\sc haz}) and P($\mbox{\sc haz} \leq -3$), which suggests that in the cases in which one has information about an un-measured metric from a measured metric in the same country-year, the measured metric is much more informative than any covariate information that allows for borrowing of strength from other countries or years.

In both tests, we also examined the validity of the main model's 95\% uncertainty intervals, checking whether 95\% of held-out values were included in the 95\% uncertainty intervals. 
Overall, the 95\% uncertainty intervals of our estimates included 92-95\% of held-out study \textsc{haz} means and prevalences in Test 1 and 95-98\% in Test 2, consistent with the expected 95\%. 
While these results are reassuring, we note that the cross-validation can only assess our quantification of predictive uncertainty in relation to the observed data. The presence of additional variability (beyond sampling variability) related to shortcomings in study quality in the nationally-representative studies makes it difficult to assess our quantification of uncertainty in the true country-level trends.  

In addition to the cross validation, we also conducted a sensitivity analysis to assess the extent to which our estimates changed in 
a model that included two additional mixture components.  
Estimates from the 7-component model were very similar to those of the 5-component model, with a median absolute difference between the estimates of mean {\sc haz} equal to 0.03 and median absolute differences of stunting and severe stunting prevalences both equal to 0.63 percentage points. Of course more rigorous methods exist for choosing a fixed value for $M$ \citep[e.g.,][]{kass:1995, Spiegelhalter:2002}, including estimation of $M$ as a model parameter \citep{richardson:1997}.

\section{Model extensions}
\label{sec:extensions}
\subsection{`Mixture of mixtures': Including data from exclusively urban and exclusively rural populations}
\label{subsec:mom}
The dataset described in \citet{NIMS:2012} excluded studies that were from selected subgroups that might have had lower or higher nutritional status than the general population.  In particular, studies of exclusively urban or exclusively rural populations were omitted from the analysis.  In this Section, we present an extension of the main model that allows the inclusion of these urban- and rural-only datapoints and that allows separate inference for each subgroup.
 
We let $s$ index `strata', with $s=u$ corresponding to an urban stratum and $s=r$ corresponding to a rural stratum.  We let $h$ index observations: In a grouped-data study that combines urban and rural individuals, there would be a single observation; in a grouped-data study that reports results separately for urban vs. rural individuals there would be two observations.  As before, $i$ indexes studies, $j$ indexes countries, and $k$ indexes regions.  The likelihood for data from stratum $s$ from study $i$ is:
$$f_{si}(y) = \sum_{m=1}^{M+1} w_{msi}\ \mathcal N\left(y|\theta_m, \sigma^2_m\right).$$

Consider a grouped-data observation $h$ that includes both urban and rural data.  Let $p_u$ be the proportion of urban dwellers  in that country-year, and let $p_r$ be the proportion of rural dwellers, such that $p_u+p_r=1$.  The resulting density for observation $h$ is a mixture of two ($M$+1)-component mixture densities.  The crux of this `mixture of mixtures' extension is that this density can be written as a simple ($M$+1)-component mixture density as in the original model:
\begin{eqnarray*}
f_h(y)  &=& p_u \sum_{m=1}^{M+1} w_{mui[h]}\ \mathcal N\left(y|\theta_m, \sigma^2_m\right)  + 
                     p_r  \sum_{m=1}^{M+1} w_{mri[h]}\ \mathcal N\left(y|\theta_m, \sigma^2_m\right) \\
&=& \sum_{m=1}^{M+1} \left(p_u w_{mui[h]} + p_r w_{mri[h]} \right)\ \mathcal N\left(y|\theta_m, \sigma^2_m\right).
\end{eqnarray*}
This trivial manipulation allows us to include data from selected subgroups without altering the fundamental structure of our model.  Furthermore, it allows us to make inference separately by stratum and to use $p_u$ and $p_r$ to combine inferences across strata and report country-level estimates as before.

We model the urban/rural effect flexibly by adding three new terms to the expression for $\alpha$.  $\gamma^c$ is a country- and component-specific stratum effect, and $\rho^c$ allows this effect to vary linearly with time (we assume no nonlinear effect for simplicity).  $c_{mi}$ is a study-$i$-specific error in the urban/rural effect.  This results in the following expanded expression for $\alpha$:
$$\alpha_{msi} = \delta^c_{mj[i]} + \varphi^c_{mj[i]}t[i]
+ u_{mj[i]t[i]} + \bm{\beta}^{\prime}_{m}\bm{x}_i 
+ a_{mi} + b_{mi}  
+ \mbox{I}_{si}\left[ \gamma^c_{mj[i]} + \rho^c_{mj[i]}t[i] + c_{mi}\right],$$
where $\mbox{I}_{si}$ is a centered stratum indicator for study $i$:
\begin{eqnarray*}
\mbox{I}_{si}= \left\{ \begin{array}{ll}
                    1 & \mbox{if stratum $s$ contains urban individuals in study $i$}\\
                    -1 & \mbox{if stratum $s$ contains rural individuals in study $i$.} \end{array} \right.
\end{eqnarray*}

The $\gamma$'s and $\rho$'s are modeled hierarchically, with the same structure as described in Sec.~\ref{subsec:hierarchical} for the $\delta$'s and $\varphi$'s:
\begin{center} \begin{tabular}{ll}
$\gamma^c_{mj} \sim t_4 \left(\gamma^r_{mk[j]}, \tau^{2\, \gamma^c}_m \right)$, & $\gamma^r_{mk} \sim \mathcal N \left(\gamma^g_{m}, \tau^{2\, \gamma^r}_m \right)$,\\
$\rho^c_{mj} \sim t_4 \left(\rho^r_{mk[j]}, \tau^{2\, \rho^c}_m \right)$, & $\rho^r_{mk} \sim \mathcal N \left(\rho^g_{m}, \tau^{2\, \rho^r}_m \right).$
\end{tabular} \end{center}
The $c$'s are modeled analogously to the $b$'s, with $c_{mi} \sim t_4(0, v^c_m)$.

An important advantage of the centered stratum indicator parameterization is that $\delta_m^c$'s can capture country-to-country differences in \textsc{haz} densities while the $\gamma_m^c$'s capture country-to-country differences in the magnitude of the difference between urban and rural populations.  In a scenario where there is a single, global effect of the urban-rural difference, $\gamma_m^g$, then $\tau_m^{\gamma^c}$ and $\tau_m^{\gamma^r}$ can both go to zero.  A non-centered formulation would have included an embedded assumption that \textsc{haz} distributions of one stratum are more variable than those of the other stratum. 

\begin{figure}
\centering
\includegraphics[width=6in]{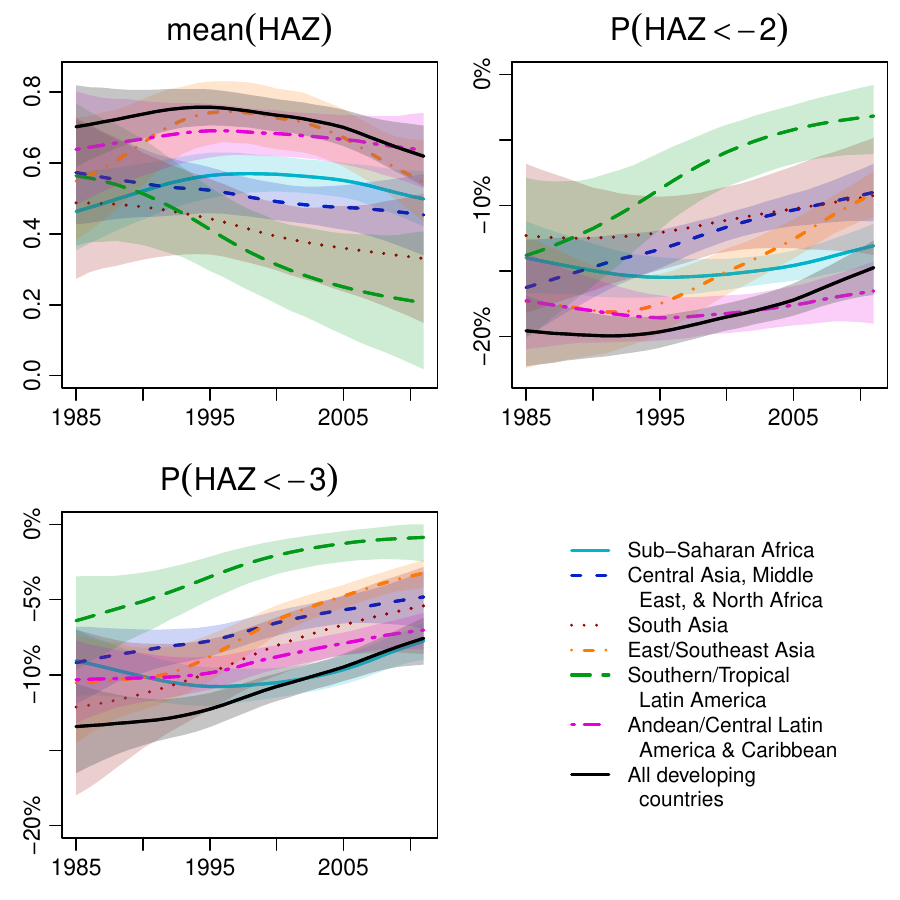}
\caption{Population-weighted region- and global-level urban-rural difference estimates with pointwise 95\% uncertainty bands. Oceania is omitted to reduce clutter and because of large posterior uncertainty.}
\label{fig:sregionFitsAlt}
\end{figure}

Based on \textsc{mcmc} samples from the posterior of the expanded model described above, Fig.~\ref{fig:sregionFitsAlt} shows trends in the urban-rural difference in mean \textsc{haz} and in the prevalence of stunting by region and for the globe as a whole. First, we note that there is an urban advantage in all regions, with Southern and Tropical Latin America showing less of a gap than other regions. In most regions, there is evidence of a narrowing of the urban-rural gap over time, with the strongest evidence and largest decreases in the gap in Southern and Tropical Latin America and (since the 1990s) East and Southeast Asia. Interestingly, the global advantage in growth for urban children is stronger than the regional advantages, reflecting the fact that regions that are more heavily urbanized tend to have higher \textsc{haz}, such that the global urban-rural difference reflects both within-region differences in growth and cross-region differences in urbanization and growth. Unlike in canonical Simpson's paradox examples, the cross-region effect is in the same direction as the within-region effects and therefore reinforces them.


This `mixture of mixtures' approach generalizes to allow inclusion of other selected subgroups.  For example, this method would allow us to treat age in a more nuanced fashion, rather than simply categorizing studies according to whether they covered the full under-5-year-old age range.  Instead, we could make age-specific estimates for each age stratum, and combine across strata via `mixture of mixtures' for studies that report \textsc{haz} in broad age bands and for the purposes of making country-level inference.

\subsection{`Sloshing': Including main-effect terms in the expression for $\alpha$}
\begin{figure}
\centering
\includegraphics[width=6in]{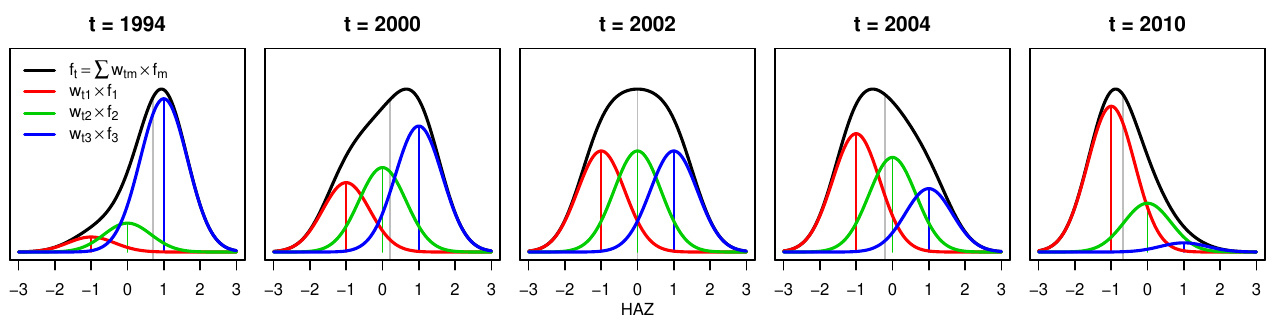}
\caption{The \textsc{haz} distribution at a given time (denoted here as $f_t$ and depicted in black), is a weighted mixture of $M+1$ (here three) normal distributions (shown in red, green, and blue). The mixture weights, $\bm{w}$, vary in time, while the locations and scales of the component normals remain constant. Here, the effect of time is linear on the probit stick-breaking scale of the $\alpha$'s. This results in a `sloshing' of the \textsc{haz} distributions, with -- for positive $\varphi$, as depicted here -- mean \textsc{haz} (shown in grey) decreasing in time.}
\label{fig:slosh}
\end{figure}

Note that each term in the expression for $\alpha_{mi}$ (\ref{eq:anova}) is mixture-component specific.  In particular, covariate effects ($\bm{\beta}_m$) and country-level intercepts ($\delta^c_{mj}$) and linear time trends ($\varphi^c_{mj}$) are each estimated separately by mixture component.  In this Section, we discuss the addition of `main effects' ($\bm{\beta}^0$, $\delta^{c0}_j$, and $\varphi^{c0}_j$, respectively) that apply across all mixture components, as in \citet{rodriguez:2009}.
Adding these three effects to the expression for $\alpha_{mi}$ yields
$$ \alpha_{mi} = \delta^{c0}_j + \delta^c_{mj[i]}+ \varphi^{c0}_j t[i] + \varphi^c_{mj[i]} t[i] + u_{mj[i]t[i]} +  \bm{\beta}^{0\prime} \bm{x}_i + \bm{\beta}^{\prime}_m \bm{x}_i + a_{mi} + b_{mi}.  $$
We note that -- though $\bm{\beta^0}$ and $\varphi^{c0}$ capture linear effects of covariates and time on the $\alpha$'s -- their effects are not linear on the scale of the $w$'s. For example, in Fig.~\ref{fig:slosh}, 
we show that the linear main effect of time ($\varphi^{c0}$) results in a `sloshing'-type movement in the \textsc{haz} distributions. For even more flexibility, one could also introduce nonlinear main effects for the $u$ terms.

We propose extending the \citet{rodriguez:2009} model to the hierarchical setting by specifying priors for the main effects $\delta^{c0}$ and $\varphi^{c0}$ that are analogous to those described in Sec.~\ref{subsec:hierarchical} for the component-specific terms, $\delta^c$ and $\varphi^c$:
\begin{center} \begin{tabular}{ll}
$\delta^{c0}_{j} \sim t_4 \left(\delta^{r0}_{k[j]}, \tau^{2\,\delta^{c0}} \right)$, & $\delta^{r0}_{k} \sim \mathcal N \left(\delta^{g0}, \tau^{2\,\delta^{r0}} \right)$,\\
$\varphi^{c0}_{j} \sim t_4 \left(\varphi^{r0}_{k[j]}, \tau^{2\,\varphi^{c0}} \right)$, & $\varphi^{r0}_{k} \sim \mathcal N \left(\varphi^{g0}, \tau^{2\,\varphi^{r0}} \right).$
\end{tabular} \end{center}
We place flat, improper priors on $\delta^{g0}$, $\varphi^{g0}$, and $\bm{\beta}^0$. We constrain the mean of each set of component-specific $\delta$'s, $\varphi$'s, and $\bm{\beta}$'s to be zero to avoid non-identifiability with their corresponding main effects.  For example, for each country $j$, we constrain $\sum_m \delta^c_{mj} = 0$ to avoid non-identifiability with $\delta^{c0}_j$.

An advantage of this specification is that the main effects parsimoniously capture variation along the `sloshing' axis (i.e. linearly in the $\alpha$'s), whereas the interaction terms provide the added flexibility needed to describe any residual variability across shapes of distributions. In cases when this flexibility is not needed, i.e., when sloshing suffices to describe all variation across shapes of distributions, the model can zero out the interaction terms.  
In our analyses to date we have not included these main effect terms, but we feel that this specification of main effects as well as the sensitivity of our modeling results to inclusion of the main effects are worth additional methodological investigation. 

\section{Discussion}
\label{sec:discussion}
Efforts to improve global health will depend on monitoring of health outcomes, and much of the monitoring will be based on incomplete data from disparate sources. Substantive interest will often focus on the distributions of health indicators, and especially on their tails.


We have specified a semiparametric model for estimating population-level distributions of these indicators.  This flexible structure allows us to estimate clinically-important tail prevalences without imposing parametric assumptions.  It also allows us to combine data sources that are aggregated in different ways and reported using a variety of metrics, accounting for redundancies across summary statistics that describe the same sample.  We borrow strength in time, covariates, and within and across regional country clusters to make inference where data are sparse or missing. The method naturally extends via our `mixture of mixtures' approach to handle additional stratification, including data that are reported separately by stratum and data that are summarized across strata.

A methodological innovation is the extension of the \citet{rodriguez:2009} model to the fully hierarchical context.  Though they refer to their model as `hierarchical', it does not have multiple nested levels, and it does not allow borrowing of strength in a data-driven manner, since the variances of all groups of effects are set to one a priori.  We generalize their model to the multi-level context and estimate all variance components.  An important consequence is that our model -- we hypothesize -- automatically favors parsimony, an especially desirable characteristic given that the model space is so large. 

Although the natural Bayesian penalty is known to favor parsimony in the context of simple models \citep{jefferys:1992}, a reviewer suggested that this belief be substantiated in the context of more complex models such as ours.  To this end, we performed a new analysis showing that at least one of our priors does indeed induce shrinkage in the context of this complex model.  In particular, we substituted flat, independent priors on each of the country-specific intercepts ($\delta^c_m$) in place of the original exchangeable shrinkage prior in order to assess the degree of shrinkage due to the prior.  (As an aside, note that in the actual analysis, we must use a shrinkage prior for the $\delta^c_m$ in order to make meaningful predictions for countries without data.  By the same token, it wouldn't have made sense to remove the shrinkage prior from the country-specific linear time slopes ($\varphi^c_m$) because without a prior, these terms are not identifiable in countries with data from fewer than two time points.)

The analysis shows that our model does indeed perform shrinkage.  This is especially true in two specific cases: (a) when we are `imputing' mean({\sc haz}) and P($\mbox{\sc haz} \leq -3$) based on data for P($\mbox{\sc haz} \leq -2$), and -- to a lesser extent -- (b) in countries where data are sparse.  Compared to the main model, in instances where only data on P($\mbox{\sc haz} \leq -2$) are available, the variance across country years in the posterior means of mean({\sc haz}) and P($\mbox{\sc haz} \leq -3$) based on the model without the shrinkage prior are 325\% and 232\% greater, respectively, than those from our model.  Similarly, but less dramatically, in countries with only 1 or 2 years of data, the variances across country-years in the posterior means of mean({\sc haz}), P($\mbox{\sc haz} \leq -2$), and P($\mbox{\sc haz} \leq -3$) are 10\%, 43\%, and 7\% greater, respectively. 

In thinking about why the variance increased so dramatically in (a), we hypothesize that although the country-specific P($\mbox{\sc haz} \leq -2$) data does inform estimation of mean({\sc haz}) and P($\mbox{\sc haz} \leq -3$) as we showed in the comparison of M1 and M2 in our cross validation -- the mean({\sc haz})  and P($\mbox{\sc haz} \leq -3$) data from neighboring countries also provide important information in the absence of direct country-level data on P($\mbox{\sc haz} \leq -2$).

So, although the literature describing shrinkage is focused on simpler models, this analysis provides evidence that shrinkage is taking place in the context of this more complex model as well, at least for the country-level intercepts.  Independent priors don't yield meaningful inference for other parameters in the model, but we hypothesize that shrinkage is happening there as well. 

\begin{figure}[!h]
\centering
\includegraphics[width=5in]{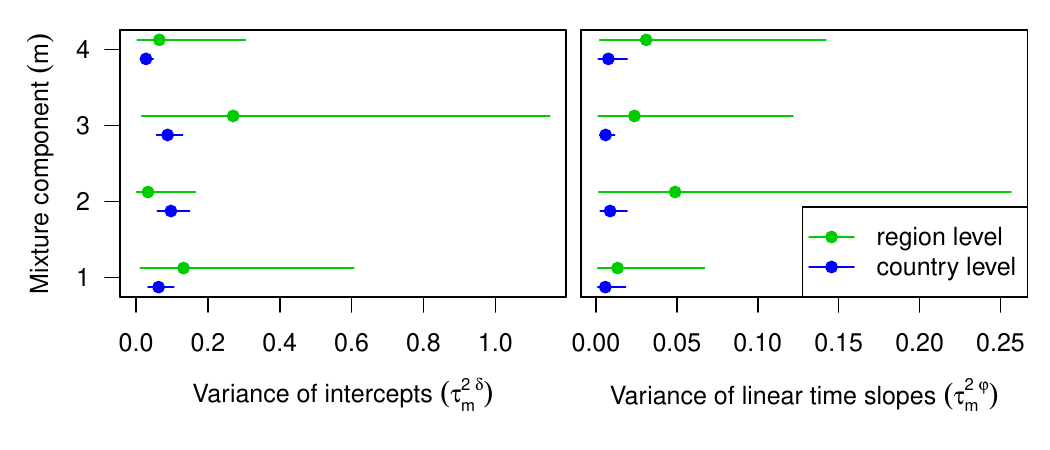}
\caption{Posterior means and 95\% CIs for the country- and region-level variances of the intercepts and linear time slopes for each of the four values of $m$.  Many of the variance components have posteriors whose CIs are bounded away from zero, suggesting that the model is not choosing to collapse down toward a simpler specification.}
\label{fig:varcomp}
\end{figure}

At the other end of the spectrum from `no shrinkage' is the possibility that our model would prefer `complete shrinkage' of one or more sets of random effects. By setting a variance component to zero, the model could shrink its associated set of random effects to zero, thereby effectively removing these effects from the model.  As seen in the results, and in  Figure \ref{fig:varcomp}, such severe shrinkage is not chosen by the model based on the fit to the data.  In particular, for the country-level intercepts, the variance components all have posteriors whose CIs are bounded away from zero, suggesting that the model is not choosing to collapse down toward a simpler specification.

A complication that we have ignored in this analysis is measurement error in children's heights, which may be especially pronounced for very young babies.  This likely leads us to overestimate the variability of the true distributions and to overestimate the tail prevalences.

While our confidence in the model is buoyed by the cross-validation results that indicate that our inference reflects the important sources of variability, there are nonetheless a number of potential model improvements that are beyond the scope of this analysis. These include consideration of additional covariates, non-linear covariate effects, and covariate interactions, including covariate effects that vary by region and time. In addition, while data sparsity led us to assume that a number of model parameters were constant across region, it would be worthwhile to investigate allowing the country-level variance components, including the autoregressive smoothing parameters, to vary by region.  


Finally, we note that in addition to analysis of height-for-age and weight-for-age \citep{NIMS:2012}, a third standard metric is weight-for-height z-score (\textsc{whz}), an important measure of short-term nutrition, including famine. Modeling outcomes such as \textsc{whz} that vary rapidly in time is difficult because real changes are likely poorly-identified relative to study design and measurement issues that can lead to large differences between surveys conducted in the same year and between one year and the next, even for large nationally-representative surveys.

\setlength{\baselineskip}{0pt}

\bibliographystyle{asa}
\bibliography{paper3}

\newpage
\setcounter{secnumdepth}{0}
\section{Appendix: Moments of the joint normal likelihood of a sample's mean, variance, and tail probabilities}

Note: for clarity, the notation in this Appendix differs slightly from that of the main body of the paper.  

We take $y_1, \cdots, y_n$ to be a sample of $n$ observations from a mixture density: $f(y) = \sum_{c=1}^C w_c \mathcal N(y|\theta_c,  \sigma^2_c)$.  For ease of notation, let  the constituent normal distributions $y_c \sim \mathcal N(\theta_c, \sigma^2_c)$ have probability density function $f_c$ and cumulative distribution function $F_c$. We begin by determining the means and variances of each of the two types of summary statistics.  We then turn our attention to the covariances across the statistics.
\begin{description} 
 \item[Sample Means:] Define $T_m = \frac{1}{n} \sum_{i=1}^n y_i$.  A property of mixture distributions is that $y$'s mean is just a weighted sum of its components' means:
\begin{eqnarray*}
\theta \stackrel{\mbox{\footnotesize def}}{=} E[y] = \sum_c w_c \theta_c.
\end{eqnarray*}
Let the variance of $y$ be denoted $\sigma^2$.  Then
\begin{eqnarray*}
\sigma^2 &\stackrel{\mbox{\footnotesize def}}{=}& V(y) = E[y^2] - \theta^2 \\
&=& \sum_c w_c(\theta^2_c + \sigma^2_c) - \theta^2,
\end{eqnarray*}
where the last equality is a property of normal mixtures.

By the central limit theorem ({\sc clt}), we have that $T_m \stackrel{\mbox{\footnotesize approx}}{\sim} \mathcal N\left(\theta, \frac{\sigma^2}{n}\right).$
\item[Sample Prevalences:] Define $T_p(x) = \frac{1}{n} \sum_{i=1}^n \mbox{I}\{y_i \leq x\}$.  Then $p_x = \sum_{c=1}^C w_cF_c(x)$, and by the {\sc clt}:
 $$T_p(x) \stackrel{\mbox{\footnotesize approx}}{\sim} \mathcal N\left(p_x, \frac{p_x(1-p_x)}{n}\right). $$

\item[Cov($T_p(x_1), T_p(x_2)$):]  Consider two cutoffs, $x_1 < x_2$.
\begin{eqnarray*}
\mbox{Cov}(T_p(x_1), T_p(x_2)) &=& \frac{1}{n^2} E\left[\sum_{i=1}^n \mbox{I}\{y_i\leq x_1\}\sum_{j=1}^n \mbox{I}\{y_j\leq x_2\}\right] - p_{x_1} p_{x_2}\\
&=& \frac{n}{n^2} E\left[\mbox{I}\{y\leq x_1\}\sum_{j=1}^n \mbox{I}\{y_j\leq x_2\}\right] - p_{x_1} p_{x_2}\\
&=& \frac{1}{n} E[\mbox{I}\{y_i\leq x_1\}\mbox{I}\{y_i\leq x_2\} +\\
& &  \hspace{.2in} (n-1)\mbox{I}\{y_i\leq x_1\}\mbox{I}\{y_j\leq x_2\}] - p_{x_1} p_{x_2}\\
&=& \frac{1}{n} (p_{x_1} + (n-1)p_{x_1}p_{x_2}) - p_{x_1} p_{x_2}\\
&=& \frac{1}{n} p_{x_1}(1-p_{x_2}).
\end{eqnarray*}
\end{description}

Calculation of the last covariance requires a bit of additional notation.  Let $\tilde \theta_{cx}$ denote the mean of the truncated $\mathcal N(y|\theta_c, \sigma^2_c) \mbox{I}\{y \leq x\}$ distribution: $$ \tilde \theta_{cx} = \theta_c - \frac{\sigma^2_c f_c(x)}{F_c(x)}.$$ 
Let $\tilde \theta_x$  denote the mean of the truncated mixture distribution, $\frac{1}{p_x}f_y \mbox{I}\{y \leq x\}$:
{\allowdisplaybreaks
\begin{eqnarray*}
\tilde \theta_x &=& \frac{1}{p_x} \int_{-\infty}^x y f_y(y)dy \\ 
&=& \frac{1}{p_x} \sum_{c=1}^C w_c \int_{-\infty}^x y f_c(y)dy \\ 
&=& \frac{1}{p_x} \sum_{c=1}^C w_c F_c(x) \int_{-\infty}^x \frac{1}{F_c(x)} y f_c(y)dy \\
&=& \frac{1}{p_x} \sum_{c=1}^C w_c F_c(x) \tilde \theta_{cx}. 
\end{eqnarray*}}
\begin{description}
\item[Cov($T_m, T_p(x)$):]    
\begin{eqnarray*}
\mbox{Cov}(T_m, T_p(x)) &=& \frac{1}{n^2} E\left[\sum_{i=1}^n y_i \sum_{j=1}^n\mbox{I}\{y_j \leq x\}\right] - \theta p_x\\
&=& \frac{n}{n^2} E\left[y_i \sum\mbox{I}\{y_j \leq x\}\right] - \theta p_x\\
&=& \frac{1}{n} E\left[y_i \mbox{I}\{y_i \leq x\} + (n-1)y_i \mbox{I}\{y_j \leq x\}\right] - \theta p_x\\
&=& \frac{1}{n} \left( E\left[y_i |y_i \leq x\right]p_x + (n-1)\theta p_x\right) - \theta p_x\\
&=& \frac{1}{n} \left( \tilde \theta_x p_x + (n-1)\theta p_x\right) - \theta p_x\\
&=& \frac{p_x}{n}(\tilde \theta_x - \theta).
\end{eqnarray*}
\end{description}
This gives us the full joint likelihood for the aggregated data:
\begin{eqnarray*}
\left(\begin{array}{c} T_m\\ T_{x_1}\\ T_{x_2} \end{array} \right) &\stackrel{\mbox{\footnotesize approx}}{\sim}&
\mathcal N\left( \left( \begin{array}{c} \theta \\ p_{x_1}\\ p_{x_2} \end{array}\right),\  
\frac{1}{n} \left( \begin{array}{ccc}
\sigma^2 & 
   &  \\
p_{x_1}(\tilde \theta_{x_1} - \theta) &  
p_{x_1}(1-p_{x_1}) & 
\\
p_{x_2}(\tilde \theta_{x_2} - \theta) &  
p_{x_1}(1-p_{x_2}) & 
p_{x_2}(1-p_{x_2})
\end{array} \right) \right). 
\end{eqnarray*}
\end{document}